\documentclass[a4paper, 12pt]{article} 
\usepackage[right=1.1in,left=1.1in,top=1.1in,bottom=1.02in]{geometry}
\usepackage{graphicx} 
\usepackage{braket}
\usepackage{mathrsfs}
\usepackage[T1]{fontenc} 
\usepackage{amsmath}
\usepackage{pxfonts}
\usepackage[T1]{fontenc}
\usepackage{amssymb}
\usepackage[round]{natbib}

\usepackage{epstopdf}
\usepackage{setspace}
\usepackage{enumitem}
\usepackage{sectsty}
\usepackage{wrapfig} 
\sectionfont{\large}
\bibpunct{(}{)}{,}{a}{}{,}
\usepackage{tikz}   
\usepackage{tikz-3dplot} 

\newenvironment{proof}[1][Proof]{\begin{trivlist}
\item[\hskip \labelsep {\bfseries #1}]}{\end{trivlist}}

\newcommand{\qed}{\nobreak \ifvmode \relax \else
      \ifdim\lastskip<1.5em \hskip-\lastskip
      \hskip1.5em plus0em minus0.5em \fi \nobreak
      \vrule height0.75em width0.5em depth0.25em\fi}
\usepackage{epigraph}

\usepackage{bbm}
\usepackage{MnSymbol}
\usepackage[all]{xy}

\usepackage{xcolor,colortbl,array,amssymb}

\makeatletter

\title{\Large{Our Fundamental Physical Space:}\\ 
\Large{An Essay on the Metaphysics of the Wave Function}} 

\author{Eddy Keming Chen\thanks{Department of Philosophy,  University of California, San Diego, 9500 Gilman Dr, La Jolla, CA 92093-0119. Website: www.eddykemingchen.net. Email: eddykemingchen@ucsd.edu  }}

\date{  \emph{The Journal of Philosophy}, 114(7), 2017: 333-65} 

\begin{document}
\bibliographystyle{plainnat}

\maketitle 

\epigraph{Already in my original paper I stressed the circumstance that I was unable to give a logical reason for the exclusion principle or to deduce it from more general assumptions. I had always the feeling, and I still have it today, that this is a deficiency.}{Wolfgang Pauli (1946 Nobel Lecture)}



\begin{abstract}

The mathematical structure of realist quantum theories has given rise to a debate about how our ordinary 3-dimensional space is related to the 3N-dimensional configuration space on which the wave function is defined. Which of the two spaces is our (more) fundamental physical space? I review the debate between 3N-Fundamentalists and 3D-Fundamentalists and evaluate it based on three criteria. I argue that when we consider which view leads to a deeper understanding of the physical world, especially given the deeper topological explanation from the unordered configurations to the Symmetrization Postulate, we have strong reasons in favor of 3D-Fundamentalism. I conclude that our evidence favors the view that our fundamental physical space in a quantum world is 3-dimensional rather than 3N-dimensional. I outline lines of future research where the evidential balance can be restored or reversed. Finally, I draw lessons from this case study to the debate about theoretical equivalence.
\end{abstract}

\hspace*{3,6mm}\textit{Keywords:}  wave function, foundations of quantum mechanics, Bohmian mechanics, GRW, Everett, fundamentality, spacetime, identical particles, Symmetrization Postulate, Manifest Image 

\newpage

\begingroup
\singlespacing
\tableofcontents
\endgroup

\vspace{30pt} 








\section*{Introduction}

\nocite{NorthSQW, MaudlinMWP, maudlin2007completeness, ney2013wave, miller2013quantum, LoewerHS, CushingFineGoldstein, forrest1988quantum, belot2012quantum, hawthorne2010metaphysician, saunders2010many, wallace2012emergent, albert1994quantum, maudlin2013nature, durr2006topological}

This is an essay about the metaphysics of quantum mechanics. In particular, it is about the metaphysics of the quantum wave function and what it says about our fundamental physical space. 

To be sure, the discussions about the metaphysics within quantum mechanics have come a long way. In the heyday of the Copenhagen Interpretation, Niels Bohr and his followers trumpeted the instrumentalist reading of quantum mechanics and the principles of complementarity,  indeterminacy, measurement recipes, and various other revisionary metaphysics. During the past three decades, largely due to the influential work of J. S. Bell, the foundations of quantum mechanics have been much clarified and freed from the Copenhagen hegemony.\footnote{For good philosophical and historical analyses about this issue, see \cite{bell2004speakable} for and \cite{cushing1994quantum}.} Currently, there are physicists, mathematicians, and philosophers of physics working on solving the measurement problem by proposing and analyzing various realist quantum theories, such as Bohmian mechanics (BM), Ghirardi-Rimini-Weber theory (GRW), and Everettian / Many-World Interpretation as well as trying to extend them to the relativistic domains with particle creation and annihilation. However, with all the conceptual and the mathematical developments in quantum mechanics (QM), its central object---the wave function---remains mysterious. 

Recently, philosophers of physics and metaphysicians have focused on this difficult issue. Having understood several clear solutions to the measurement problem, they can clearly formulate their questions about the wave function and their disagreements about its nature. Roughly speaking, there are those who take the wave function to represent some mind-dependent thing, such as our ignorance; there are also people who take it to represent some mind-independent reality, such as a physical object, a physical field, or a law of nature. 

In this paper, I shall assume realism about the wave function---that is, what the wave function represents is something real, objective, and physical. (Thus, it is not purely epistemic or subjective).\footnote{I take it that there are good reasons to be a realist about the wave function in this sense. One reason is its important role in the quantum dynamics. See \cite{AlbertEQM} and \cite{NeySOTDQU}.  The recently proven PBR theorem lends further support for this position. See the original PBR paper \cite{pusey2012reality} and Matthew Leifer's excellent review article \cite{leifer2014quantum}.}
I shall conduct the discussion  in non-relativistic quantum mechanics, noting that the lessons we learn here may well apply to the relativistic domain.\footnote{See \cite{pittphilsci11117} for an insightful discussion on the complications when we transfer the discussion to the relativistic domain with particle creation and annihiliation. I take it that a clear ontology of QFT will provide a natural arena for conducting a similar debate. Since a consistent QFT is still an incomplete work-in-progress (although an effective quantum field theory is highly useful and predictively accurate, and there have been proposals with clearer ontology such as Bell-type QFT), I believe that there are many values in conducting the discussion as below---in the non-relativistic QM---although we know that it is only an approximation to the final theory. The  clarity of the non-relativistic QM, if for no other purposes, makes many issues more transparent.}  To disentangle the debate from some unfortunately misleading terminologies, I shall adopt the following convention.\footnote{I am indebted to \cite{maudlin2013nature} for the following distinctions.}
I use \emph{the quantum state} to refer to the physical object and reserve the definite description \emph{the wave function} for its mathematical representation, $\Psi$.  In the position representation (which I use throughout this paper), the domain of the wave function is all the ways that fundamental particles can be arranged in space and the codomain is the complex field $ \mathbb{C}$.\footnote{It is reasonable to wonder whether our definition applies to ``flashy'' GRW (GRWf) or mass-density versions of GRW (GRWm). Although particles are not fundamental in these theories, the definitions of GRWf and GRWm and the mathematical structures of these theories can be captured by using wave functions defined on the particle-location configuration space.  \emph{When interpreting the physical theory}  (either according to the low-dimensional view or the high-dimensional view described below), if we think of the wave function as representing something \emph{in addition to} the material ontology, we can regard the wave function space as having a structure that is independent from the material ontology (and the configurations of material stuffs). We can call the domain of the wave function ``the configuration space,'' but we do not need to understand it  literally as the space of configurations of the material ontology. How the material ontology is connected to the wave function on ``the configuration space'' is explained by the definitions (such as the mass-density function $m(x,t)$ in GRWm) and the dynamical laws. Interpreted this way, many  arguments that rely on the configuration space of particle locations,  including those in Sections 2 and 3, apply directly to GRWf and GRWm. The exceptions are those arguments that rely on a \emph{justification} or an \emph{explanation} of the configuration space. Therefore, our arguments in Section 4.2 apply most smoothly to theories with a particle ontology such as Bohmian Mechanics and much less smoothly to GRW theories.} 
(For convenience and simplicity, I leave out spin when discussing the value of the wave function, noting that it can be added as additional degrees of freedom (spinor values). Hence, $\Psi \in L^{2}(\mathbb{R}^{3N}, \mathbb{C}).$) The domain of the wave function is often called \emph{the configuration space}, and it is usually represented by $\mathbb{R}^{3N}$, whose dimension is 3 times N, where N is the total number of the fundamental particles in the universe. (I shall represent the configuration space by $\mathbb{R}^{3N}$ instead of $\mathbb{E}^{3N}$, noting that the latter would be more physical than the former.) For the wave function of a two-particle system in $\mathbb{R}^{3}$, the configuration space is $\mathbb{R}^{6}$. For the wave function of our universe, whose total number of fundamental particles most likely exceeds $10^{80}$, the configuration space has at least as many dimensions as $10^{80}$.

Our key question is: given realism about the quantum state, how is the configuration space related to our familiar 3-dimensional space? There are, roughly speaking, two views corresponding to two ways to answer that question.\footnote{I leave out the middle position between the two views, which says that both the 3N space and the 3D space are fundamental. See \cite{DorrFOOSQW} for an exploration of that view.}

\begin{description}
\item[3N-Fundamentalism] $\mathbb{R}^{3N}$ represents the fundamental physical space in quantum mechanics. Our ordinary space, represented by $\mathbb{R}^{3}$, is less fundamental than  and stands in some grounding relations (such as emergence) to $\mathbb{R}^{3N}$.\footnote{This view is often called Wave Function Realism, although we should remember that this is unfortunately misleading terminology, for its opponents are also realists about the wave function. Under my classification, 3N-Fundamentalists include David Albert, Barry Loewer, Alyssa Ney, Jill North, and \emph{arguably}, the time-slice of J. S. Bell that wrote ``Quantum Mechanics for Cosmologists:'' ``\emph{No one can understand this theory until he is willing to think of $\psi$ as a real objective field rather than just a `probability amplitude'. Even though it propagates not in 3-space but in 3N-space.}'' (Bell (1987) p.128, his emphasis.)} 
\end{description}

\begin{description}
\item[3D-Fundamentalism] $\mathbb{R}^{3}$ represents the fundamental physical space in quantum mechanics. The configuration space, represented by $\mathbb{R}^{3N}$, is less fundamental than and stands in some grounding relations (such as mathematical construction) to $\mathbb{R}^{3}$.\footnote{This view is often taken to be a commitment to Primitive Ontology or Primary Ontology. Under my classification, 3D-Fundamentalists include Bradley Monton, Detlef D\"urr, Sheldon Goldstein, Nino Zangh\`i, Roderich Tumulka, James Taylor, Ward Struvye, Valia Allori, Tim Maudlin, Michael Esfeld, CianCarlo Ghirardi, and Angelo Bassi. Depending on how one thinks about the nature of the wave function, there are three ways to flesh out 3D-Fundamentalism: (1) the wave function is a physical object---a multi-field (more later); (2) the wave function is nomological or quasi-nomological; (3) the wave function is a \emph{sui generis} entity in its own category. No matter how one fleshes out the view, the following discussion should be relevant to all, although (1) is the most natural viewpoint to conduct the discussion. In a companion paper, I discuss the viability of (2) in connection to David Lewis's thesis of Humean supervenience. In a future paper, I plan to discuss the viability of (3). In any case, defending (2) and (3) takes much more work and might in the end decrease the plausibility of  3D-Fundamentalism. If the following discussion is right, since (1) is the most plausible competitor with 3N-Fundamentalism with a high-dimensional field, it has been premature for 3D-Fundamentalists to give up the defense of (1). Indeed, a careful examination of the pros and the cons reveals its superiority over 3N-Fundamentalism (and its internal competitors (2) and (3)). I shall not defend this claim here, and for this paper the reader does not need to take sides in this internal debate among 3D-Fundamentalists.}
\end{description}
3N-Fundamentalism and 3D-Fundamentalism are the targets of the recent debates under the name ``The Metaphysics of the Wave Function,'' ``Wave Function Realism versus Primitive Ontology,'' and ``Configuration Space Realism versus Local Beables.'' (To be sure, these debates often branch into discussions about other important issues, such as the nature of the quantum state: whether $\Psi$ is a field, a law, or a \emph{sui generis} physical object.) In those debates, the most common considerations include: (1) the dynamical structure of the quantum theory, and (2) the role of our ordinary experiences. Usually, the opposing thinkers are taken to assign different weights to  these considerations. 

For example, Jill North (\citeyear{NorthSQW}) summarizes the disagreement as follows: 
\begin{quote}
This brings us to a basic disagreement between wave function space and ordinary space views: how much to emphasize the dynamics in figuring out the fundamental nature of the world. Three-space views prioritize our evidence from ordinary experience, claiming that the world appears three-dimensional because it is fundamentally three-dimensional. Wave function space views prioritize our inferences from the dynamics, claiming that the world is fundamentally high-dimensional because the dynamical laws indicate that it is.\footnote{\cite{NorthSQW}, p. 196.}
\end{quote}


However, as I shall argue, the common arguments based on (1) and (2) are either unsound or incomplete. Completing the arguments, it seems to me, render the overall considerations based on (1) and (2) \emph{roughly} in favor of 3D-Fundamentalism. However, there is another relevant consideration: (3) the deep explanations of striking phenomena. Here, we find a more decisive argument in favor of 3D-Fundamentalism as it leads to a deeper understanding of the mathematical symmetries in the wave function, known as the Symmetrization Postulate. Since 3D-Fundamentalism explains the Symmetrization Postulate much better than 3N-Fundamentalism does, we have strong reasons to accept 3D-Fundamentalism. 

I shall argue that the considerations (1), (2), and (3), taken together, favors 3D-Fundamentalism over 3N-Fundamentalism. Here is the roadmap. 
In \S 1, I shall argue, against the 3N-Fundamentalist, that the common argument from the dynamical structure of quantum mechanics is unsound, and that after proper refinement it no longer provides strong reason in favor of 3N-Fundamentalism. (As a bonus, I will provide a direct answer to David Albert's question to the primitive ontologists: what is their criterion for something to be the fundamental physical space?)
In \S 2, I shall argue, against the typical 3D-Fundamentalist, that the common argument from the Manifest Image can be resisted with spatial functionalism, but such a response can be weakened by examining the details of the functionalist strategies. This suggests that the argument from the Manifest Image, though no longer decisive, still has some force. 
In \S 3, I shall explore which position leads to a better understanding of the physical world and formulate a new argument based on the recent mathematical results about the relationship between identical particles and the symmetries in the wave function. The deeper topological explanation offered by 3D-Fundamentalism strongly suggests that our fundamental physical space in a quantum world is 3-dimensional rather than 3N-dimensional.

Finally, I shall offer future directions of research,  point out how a 3N-Fundamentalist might be able to respond to the last two considerations, and offer additional explanations to restore or reverse the evidential balance between  3D-Fundamentalism and  3N-Fundamentalism. 

As for the scope, this essay is more or less orthogonal to the debate between 3D and 4D views about time and persistence. However, the analysis in this essay should be of interest to participants in those debates as well as to philosophers working on the nature of the quantum state, the fundamentality / emergence of space-time, and mathematical explanation in physics. As we shall see, our case study also motivates a particular way of understanding the equivalence of theories in terms of explanatory equivalence.


\section{Evidence \#1: The Dynamical Structure of Quantum Mechanics}

At first glance, 3N-Fundamentalism looks like a highly surprising idea. What can possibly motivate such a revisionary metaphysical thesis? The initial motivation was the reality of the non-local and non-separable quantum state and the phenomena of quantum entanglement---causal influences that are unweakened by arbitrary spatial separation. Perhaps what we see in front of us (the interference patterns on the screen during the double-slit experiment), the idea goes, are merely projections from something like a ``Platonic Heaven''---some higher-dimensional reality on which the dynamics is perfectly local. 

The initial motivation leads to our first piece of evidence in the investigation of our fundamental physical space. Such evidence is based on the dynamical laws and the dynamical objects in quantum mechanics. This seems to be the strongest evidence in favor of 3N-Fundamentalism. In this section, I suggest that the dynamical structure of the quantum theory, under closer examination, does not support the view that the fundamental physical space in quantum mechanics is 3N-dimensional. 

\subsection{The Argument for 3N-Fundametalism}

Here is one way to write down the argument for 3N-Fundamentalism based on considerations of the dynamics.\footnote{Strictly speaking, the following are based on considerations of both the kinematics and the dynamics of the quantum theory. Nevertheless, the kinematical structure of the quantum wave function plays an important role in determining the dynamics of the material ontology (Bohmian particles, mass densities, etc) that move in the physical space.} 

\begin{description}
\item[P1] We have (defeasible) reasons to infer a fundamental structure of the world that matches the dynamical structure of our fundamental physical theory. [The Dynamical Matching Principle]
\item[P2] The dynamical structure of quantum mechanics (a candidate fundamental physical theory) includes the quantum state which is defined over a 3N-dimensional configuration space. [The 3N-Structure of the Quantum State]

\bigskip\hrule\smallskip

\item[C1] We have (defeasible) reasons to infer that our fundamental physical space is 3N-dimensional. 
\end{description}

\textbf{P1}---the Dynamical Matching Principle---follows from a more general principle of inference from physical theories to metaphysical theories:
\begin{description}
\item[The Matching Principle] We have (defeasible) reasons to infer a fundamental structure of the world that matches the structure of our fundamental physical theory. We should infer no more and no less fundamental structure of the world than what is needed to support the structure of our fundamental physical theory. 
\end{description}
The Matching Principle can be a highly useful guide for empirically-minded metaphysicians.\footnote{This principle comes from \cite{NorthSQW} and \cite{NorthST} (forthcoming). See also \cite{AlbertEQM} and \cite{albert2015after} for ideas in a similar spirit.} For example, we can use it to infer that the classical space-time is  Galilean rather than Newtonian. This inference  agrees with our intuitions. Moreover, the inference is not based on controversial empiricist (perhaps verificationalist) assumptions about eliminating unobservable or unverifiable structure.  

\subsection{The Assumptions in Premise 2}

Suppose that something like the Matching Principle is true and the particular application with the Dynamical Matching Principle is justified. Then the argument would rest on \textbf{P2}. However, does the quantum state, represented by the wave function $\Psi$, really have a 3N-dimensional structure? In other words, does the wave function have to be defined over a high-dimensional configuration space? 

Many people would answer positively. How else can we define the wave function, if not on a high-dimensional configuration space? For surely the wave function has to take into account quantum systems with entangled subsystems that have non-local influences on each other, which have to be represented by wave functions that are non-separable. To take these into account, the usual reasoning goes, the wave function of an N-particle system (say, our universe with $10^{80}$  particles), in general, has to be defined as a field over on a 3N-dimensional space. Alyssa Ney (\citeyear{NeySOTDQU}), for example, endorses this line of reasoning:

\begin{quote}
...[Entangled] states can only be distinguished, and hence completely characterized in a higher-than-3-dimensional configuration space. They are states of something that can only be adequately characterized as inhabiting this higher-dimensional space. This is the quantum wavefunction.\footnote{\cite{NeySOTDQU}, p. 556.}
\end{quote}

North's reasoning is similar:
\begin{quote}
In quantum mechanics, however, we must formulate the dynamics on a high-dimensional space. This is because quantum mechanical systems can be in entangled states, for which the wave function is nonseparable. Such a wave function cannot be broken down into individual three-dimensional wave functions, corresponding to what we think of as particles in three-dimensional space. That would leave out information about correlations among different parts of the system, correlations that have experimentally observed effects. Only the entire wave function, defined over the entire high-dimensional space, contains all the information that factors into the future evolution of quantum mechanical systems.\footnote{\cite{NorthSQW}, p. 190}
\end{quote}
There are two  assumptions that are worth making explicit in the above reasonings (to be sure, North and Ney discuss them in their respective papers):
\begin{description}
\item[$\Psi$-as-a-Field]  The quantum state, represented by the wave function $\Psi$, is a field. 
\item[Field-Value-on-Points]  The space that a field ``lives on'' is the smallest space where it assigns a value for each point in that space. 
\end{description}
The two assumptions\footnote{The qualification ``smallest'' is added to ensure that the principles render a unique space that a field lives on.} are explicit in David Albert's argument in his important paper ``Elementary Quantum Metaphysics'': 
\begin{quote}
The sorts of physical objects that wave functions are, on this way of thinking, \emph{are} (plainly) fields---which is to say that they are the sorts of objects whose states one specifies by specifying the values of some set of numbers at every point in the space where they live, the sorts of objects whose states one specifies (in this case) by specifying the values of two numbers (one of which is usually referred to as an amplitude, and the other as a phase) at every point in the universe's so-called configuration space.\footnote{ \cite{AlbertEQM}, p. 278}
\end{quote}
However, as I shall argue, there are good reasons to reject both assumptions.

To deny \textbf{$\Psi$-as-a-Field}, one might instead propose that $\Psi$ is not strictly speaking a field (in any classical sense). For example, an electromagnetic field assigns a  field  value to every point in the 3-dimensional space, and the values are meaningful partly because multiplying the field by a constant will result in a different field and different dynamics. (The gauge freedom in the electric potential therefore indicates its relation to something more physical and more fundamental, such as the electric field.\footnote{Thanks to Michael Townsen Hicks for alerting me to the connection.}) However, the quantum wave function, multiplied by a global phase factor ($\Psi \Rightarrow  e^{i\theta}\Psi$), remains physically the same.  Insisting on \textbf{$\Psi$-as-a-Field} will lead one to recognize physically indistinguishable features (that play no additional explanatory role) as physically real and meaningful. It is desirable, therefore, that we do not recognize $\Psi$ as a fundamental field.\footnote{However, it is possible to provide a gauge-free account of the wave function by giving a nominalistic and intrinsic account of quantum mechanics. In a future paper, I will provide the beginning of such an account. But the next point  holds regardless of the success of that program.} 

However, even if one were to embrace the indistinguishable features brought about by \textbf{$\Psi$-as-a-Field}, one can still deny \textbf{Field-Value-on-Points}. We usually think, as Albert describes, that the fundamental physical space is the fundamental space that the dynamical object ``lives on.'' In particular, the fundamental space that a field lives on is one in which we specify the state of the field at a time by specifying the values at \emph{each point}. This  is indeed the case for the  electromagnetic field. But why think that this is an essential feature for a physical field? 

Suppose that a fundamental field is mathematically represented by a function from one space to another space. Such a function can be as general as one likes. A special case, of course, is the electromagnetic  field that assigns values to each point in $\mathbb{R}^{3}$. But another one can be a quantum multi-field that assigns values to every N-tuple of points in $\mathbb{R}^{3}$. That is, the function takes N arguments from $\mathbb{R}^{3}$.  If we think of the  electromagnetic field as assigning properties to each  point in $\mathbb{R}^{3}$, we can think of the quantum multi-field as assigning properties to each plurality of N points in $\mathbb{R}^{3}$ (N-plurality). The multi-field defined over N-pluralities of points in $\mathbb{R}^{3}$ is, unsurprisingly, equivalent to the quantum wave function defined over each point in  $\mathbb{R}^{3N}$. \footnote{This approach of taking the wave function as a ``multi-field'' is explained in \cite{forrest1988quantum}  Ch. 6 and \cite{belot2012quantum}. I have heard that for many working mathematical physicists in the Bohmian tradition, this has always been how they think about the wavefunction.}

However, this approach has a disadvantage.  On this conception of the multi-field, we lose Albert's very clear criterion for something to be the fundamental physical space (based on the space that the fundamental field ``lives on''). Indeed, Albert asks,\footnote{Albert raised this question during his talk ``On Primitive Ontology'' at the 2014 Black Forest Summer School in Philosophy of Physics.} having given up his criterion, what would be an alternative conception of the fundamental physical space that the 3D-Fundamentalists can give? 

No one, as far as I know, has provided an alternative criterion for something to be the fundamental physical space. But would it be terrible if there were none? Some people would not worry, for the following reason. When we work on the metaphysics of physical theories, we have in mind complete physical theories---theories that specify their fundamental physical spaces; that is, for each theory, we do not need to infer its fundamental space from the other parts of the theory. For each complete theory, the fundamental space is given. Our task is to decide which one (among the complete theories) is the best theory, by considering their theoretical virtues and vices (especially their super-empirical virtues such as simplicity, parsimony, explanatory depth, and so on).  Albert's question is not an urgent one, because it is about a different feature of  theories: how we arrive at them in the first place. We could wonder whether we have inferred in the right way when we say that the fundamental space of the theory is 3-dimensional. But we could also just ask which theory is more plausible:  the quantum theory defined on $\mathbb{R}^3$ or the quantum theory defined on $\mathbb{R}^{3N}$.

However, we might construe Albert's question in a different way, as asking about a distinctive theoretical virtue, namely, internal coherence. In contrast to the previous construal, this perspective would make Albert's worry highly relevant to the interpretive projects in the metaphysics of quantum mechanics.  As I think of it, internal coherence is distinct from logical consistency. A metaphysical / physical theory can be logically consistent without being internally coherent, in the sense that it might contain theoretical parts that are in serious tension (though no logical incompatibility) with each other. What the tension is can be precisified according to the types of the theory. For example, a theory with a field-like entity defined not on individual points but on non-trivial regions in the fundamental space has theoretical parts that are in tension. Hence, one way for a theory to lack internal coherence is for it to fail Albert's criterion of  \textbf{Field-Value-on-Points}.

Construed in that way, \textbf{Field-Value-on-Points} is a precisification of the theoretical virtue of internal coherence. However, I shall argue that Albert's argument against 3D-Fundamentalism will not work even if internal coherence is a theoretical virtue, for there is another equally valid way to precisify internal coherence. 

Let us consider the following proposal for a first pass:
\begin{description}
\item[Field-Value-on-Pluralities]  The space that a field ``lives on'' is the smallest space where it assigns values for  pluralities of points in that space. 
\end{description}
The modified criterion no longer requires the field to take on values at each point in the space. But this seems to lead to another disastrous consequence for the 3D-Fundamentalist who holds the multi-field view, for the smallest space where it assigns values for a plurality of points is not $\mathbb{R}^{3}$ but $\mathbb{R}^{1}$ (a linear subspace of $\mathbb{R}^{3}$) where 3N  fundamental particles move around (in a 1-dimensional space).\footnote{Thanks to David Albert for pointing out this worry.} Should the 3D-Fundamentalist, based on the above criterion, accept  $\mathbb{R}^{1}$ as the fundamental physical space? 

That initial worry dissolves itself if we take the particles to be indistinguishable and physically identical, and we take the configuration space to be an unordered configuration space (we will return to this proposal in \S 3). If the fundamental physical space is $\mathbb{R}^{1}$, then two particles cannot pass through each other (see diagram below). For two particles to pass through each other (which we would like to make possible given the degrees of freedom of the particles in the 3D space and the complexity of our world), there must be a moment in time where they occupy the same point, which means that the wave function will be undefined at that moment. (Recall that we represent an unordered configuration as a set instead of an ordered tuple, and that the wave function is defined as functions from sets of N points in  $\mathbb{R}^{3}$ to complex numbers. At the instant where the two particles occupy the same point (as the following diagram in $\mathbb{R}^1$ shows), the configuration will lose at least one point, dropping down to a configuration of 3N-1 points. The wave function would therefore be undefined.) However, things would be different as soon as we go to a bigger space, say, $\mathbb{R}^{2}$. A particle can ``pass through'' another by missing it by a small distance (such as particle 2* in the second diagram below) or going around it, perhaps in a non-straight line, with arbitrarily small positive distance $\epsilon$ (such as particle 2).\footnote{Thanks to Sheldon Goldstein and Nino Zangh\`i for pointing this out to me.} Thus, the bigger space $\mathbb{R}^{2}$ provides more ``expressive power'' than $\mathbb{R}^{1}$. \\ \\

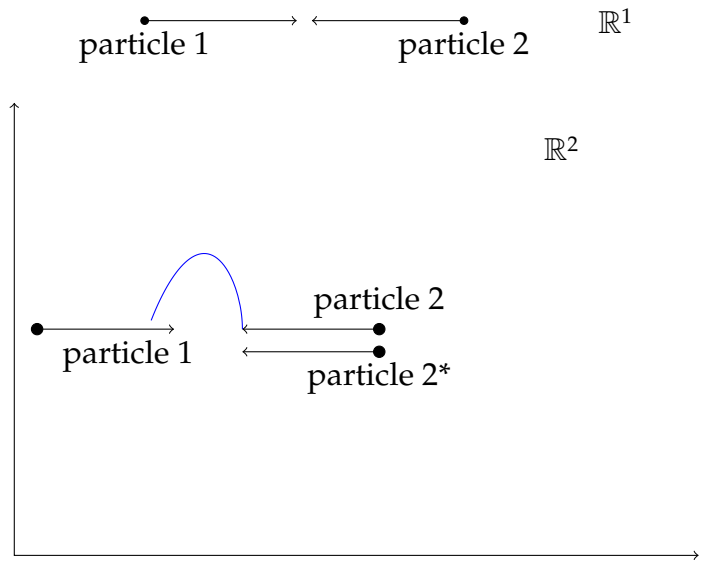
\centerline{
\begin{tikzpicture}
\draw [->] (0,0) -- (2,0);
\draw [<-] (2.2, 0) -- (4.2, 0); 
\draw[fill] (0 , 0) circle [radius=0.05];
\node [below] at (0, 0) {particle 1};
\draw[fill] (4.2 , 0) circle [radius=0.05];
\node [below] at (4.2, 0) {particle 2};
\node at (6.2, 0) {$\mathbb{R}^1$};
\end{tikzpicture}}
\bigskip

\centerline{
\begin{tikzpicture}[scale=3]
\draw [<->] (0,2) -- (0,0) -- (3,0); 
\draw [->] (0.1,1) -- (0.7,1);
\draw [<-] (1,1) -- (1.6,1);
\draw [<-] (1,0.9) -- (1.6,0.9);
\draw[fill] (0.1 , 1) circle [radius=0.025];
\draw[fill] (1.6 , 1) circle [radius=0.025];
\draw[fill] (1.6 , 0.9) circle [radius=0.025];
\node [below] at (0.5, 1) {particle 1};
\node [above] at (1.6, 1) {particle 2};
\node [below] at (1.6, 0.9) {particle 2*};
\node at (2.4,1.8) {$\mathbb{R}^2$};
\draw[blue] (0.6, 1.0385) --
(0.61, 1.06372) -- (0.62, 1.08756) -- (0.63, 1.11012) -- (0.64,
1.13147) -- (0.65, 1.15166) -- (0.66, 1.17074) -- (0.67, 1.18874) -- (0.68,
1.20568) -- (0.69, 1.22157) -- (0.7, 1.23643) -- (0.71, 1.25026) -- (0.72,
1.26307) -- (0.73, 1.27486) -- (0.74, 1.28561) -- (0.75, 1.29534) -- (0.76,
1.30402) -- (0.77, 1.31165) -- (0.78, 1.31821) -- (0.79, 1.32369) -- (0.8,
1.32806) -- (0.81, 1.33131) -- (0.82, 1.3334) -- (0.83, 1.33431) -- (0.84,
1.334) -- (0.85, 1.33244) -- (0.86, 1.32956) -- (0.87, 1.32533) -- (0.88,
1.31966) -- (0.89, 1.3125) -- (0.9, 1.30373) -- (0.91, 1.29325) -- (0.92,
1.2809) -- (0.93, 1.26649) -- (0.94, 1.24976) -- (0.95, 1.23032) -- (0.96,
1.2076) -- (0.97, 1.18065) -- (0.98, 1.14763) -- (0.99, 1.1038) -- (0.991,
1.09836) -- (0.992, 1.09261) -- (0.993, 1.0865) -- (0.994, 1.07994) -- (0.995,
1.07282) -- (0.996, 1.06497) -- (0.997, 1.0561) -- (0.998, 1.04563) -- (0.999,
1.03209) -- (0.9991, 1.03042) -- (0.9992, 1.02866) -- (0.9993,
1.02679) -- (0.9994, 1.02478) -- (0.9995, 1.0226) -- (0.9996, 1.02019) -- (0.9997,
1.01747) -- (0.9998, 1.01424) -- (0.9999, 1.01005) -- (0.9999,
1.01005) -- (0.99991, 1.00953) -- (0.99992, 1.00898) -- (0.99993,
1.0084) -- (0.99994, 1.00778) -- (0.99995, 1.0071) -- (0.99996,
1.00634) -- (0.99997, 1.00549) -- (0.99998, 1.00448) -- (0.99999, 1.00317) -- (1,
1) ;
\end{tikzpicture}} 
\medskip

Why not, then, take  $\mathbb{R}^{2}$ instead of $\mathbb{R}^{3}$ as our fundamental physical space, since it is also a linear subspace of $\mathbb{R}^{3}$? There are two reasons. First, we do not know whether the total number of apparent particles in 3D space is even, which is required if the fundamental dynamics is given by 3N/2 many particles in $\mathbb{R}^{2}$. However, we do know (as the reader can easily verify) that the  number of fundamental particles times the dimensions of the fundamental space is a multiple of 3. It seems plausible that, at least in the quantum theory, the fundamental physical space should remain neutral about whether there are $2k$ or $2k-1$ particles. Second, even more than the dynamics of 3N particles in $\mathbb{R}^{1}$, the dynamics of 3N/2 many particles, especially the Hamiltonian function, (even if we assume N is even) is unlikely to be natural.\footnote{There may be many other reasons why the 3-dimensional space is special. In \S 3.2, we consider the case of identical particles, for which spaces with fewer than 3 dimensions would allow fractional statistics that correspond to anyons, in addition to fermions and bosons.}  
It is appropriate to add a neutrality constraint and a naturality constraint to the previous criterion of the fundamental physical space: 
\begin{description}
\item[Field-Value-on-Pluralities*]  The space that a field ``lives on'' is the smallest and most natural space that is neutral to the parity of the total number of fundamental particles and that the field assigns values for pluralities of points in that space. 
\end{description}
If we regard Albert's question as tracking the theoretical virtue of what I call ``internal coherence,'' then I take Field-Value-on-Pluralities*  as an equally attractive way of precisifying internal coherence. Given this precisification, 3D-Fundamentalism wins, as $\mathbb{R}^{3}$ satisfies this criterion and is the space where the quantum multi-field lives on. Without offering strong reasons to favor his precisification, Albert's argument  falls short of delivering the conclusion.

In this section, I have examined and improved the Argument against 3D-Fundamentalism and still found its key premise unjustified. Therefore, our first piece of evidence from the dynamical structure of quantum mechanics underdetermines our choice between 3N-Fundamentalism and 3D-Fundamentalism.




\section{Evidence \#2: Our Ordinary Perceptual Experiences}

In this debate, many people have pointed out the obvious---our Manifest Image---the tables, chairs, and experimental devices in front of us. How can these apparently 3-dimensional objects exist if reality is vastly high-dimensional? If the high-dimensional quantum mechanics could not explain the apparent reality of pointer readings in 3 dimensions, from which we come to be justified in believing in quantum mechanics, the theory would surely be self-undermining. 

Hence it is common for 3D-Fundamentalists to use the second piece of evidence---our ordinary perceptual experiences---to argue in favor of their view. In this section, I suggest that although the evidence seems to support their view over the alternative---3N-Fundamentalism, the evidential force depends crucially on the fate of functionalism. 

\subsection{The Argument for 3D-Fundamentalism}

Here is one way to write down the argument for 3D-Fundamentalism based on considerations of the Manifest Image of 3-dimensional observers and pointers. (I shall use relatively weak assumptions to generate a relatively strong argument.)

\begin{description}
\item[P3] If we cannot locate the Manifest Image of human observers and pointer readings in an empirical (fundamental) scientific theory, then we have (defeasible) reasons  against that theory. 
\item[P4] We cannot locate the Manifest Image of human observers and pointer readings in the quantum theory with 3N-Fundamentalism, which is an empirical (fundamental) scientific theory. 
\bigskip\hrule\smallskip

\item[C2]  We have (defeasible) reasons against the quantum theory on 3N-Fundamentalism. 
\end{description}
The first premise---\textbf{P3}---does not depend on an insistence of respecting the common sense or the reality of the Manifest Image.  Instead,  it follows from a more general principle: 
\begin{description}
\item[Self-Undermining] If we arrive at an empirical (fundamental) theory T based solely on evidence E, and either we see that T entails that E is false without providing a plausible error theory, or we see that T entails that our acquiring E is produced by an unreliable process, or we see an ineliminable explanatory gap from T to the truth of E (or some nearby propositions), then we have (defeasible) reasons against  T. 
\end{description}
To see how \textbf{P3}  follows from \textbf{Self-Undermining}, we need to rewrite \textbf{P3} in more precise terms. 
\begin{description}
\item[P3*] If an empirical (fundamental) scientific theory contains an ineliminable explanatory gap for the true claims (or some nearby propositions) about human observers' pointer readings after experiments, then we have (defeasible) reasons against that theory. 
\end{description}

\textbf{P3*} seems highly plausible.\footnote{This is similar to the demand for ``empirical coherence'' in \cite{barrett1999quantum} and \cite{healey2002can}. The principle, I take it, is weaker than the often cited principle in \cite{maudlin2007completeness} that the physical theory makes contact with evidence via ``local beables.'' \cite{huggett2013emergent} discuss, in light of the recent developments of quantum gravity and string theory, the complications of that requirement.} The key idea is that our belief in a fundamental scientific theory is not justified in a void, or properly basic, or merely coherent with respect to the rest of our metaphysical beliefs. Rather, we are justified in believing in it because we (or the scientific community as a whole) have acquired sufficient empirical evidence through the pointer readings connected to particular experimental set-ups. These pointer readings can be in principle reduced to the positions of entities in a 3-dimensional space. And we acquire such crucial pieces of evidence through perception and testimony. Each person in the society may not acquire the evidence in a direct way.  However, there has to be some direct perception of the 3-dimensional recordings of the positions that justifies our beliefs in the scientific theory. That is, we come to believe in the scientific theory on the basis of our observations about the arrangement of macroscopic objects in the 3-dimensional space. Now, if the scientific theory cannot explain the truth of such observations of pointer readings, or if it turns out that we would be systematically wrong about the arrangements of these macroscopic objects in the 3-dimensional space, then we acquire an undermining defeater: the theory suggests that our evidence for the theory is false. This is especially objectionable for scientific theories that attempt to give a comprehensive explanation for the empirical world that  includes the human observers, pointers, and laboratory set-ups.

\subsection{The Assumptions in Premise 4}

If we accept \textbf{P3*}, then the success of the argument depends on the second premise---\textbf{P4}. \textbf{P4} is the claim that we cannot find the Manifest Image of human observers and pointer readings in the quantum theory on 3N-Fundamentalism. In other words, there is  an ineliminable explanatory gap for the true claims (or some nearby propositions) about human observers' position readings of the experimental set-ups. 

Many people sympathetic to 3D-Fundamentalism take this premise to be straight-forward.\footnote{See, for example, \cite{allori2013primitive} and \cite{maudlin2013nature}.}
However,  3N-Fundamentalists have offered a proposal about how to close the explanatory gap. Their general strategy goes by the name of ``functionalism.''\footnote{It is different from functionalism in philosophy of mind, as it is concerned not with minds but macroscopic physical objects, but the two theories have interesting similarities that are worth exploring further.}  


In our context here, ``functionalism'' refers to a formulation of the sufficient condition for what counts as emergent and real objects:

\begin{description}
\item[Functionalism] If we have a fundamental theory of the world in which we can define a mapping from the fundamental degrees of freedom to $Q_1, Q_2, ... Q_n$, and $Q_1, Q_2, ... Q_n$ have the same counterfactual profile of what we take to be the constituents of the Manifest Image, then $Q_1, Q_2, ... Q_n$ are real, emergent entities from that fundamental theory and our discourse  about the Manifest Image is made true by the fundamental theory via the functionalist mapping.\footnote{Thanks to an anonymous referee, I realize that this definition of functionalism leaves open the question how to understand the notion of ``counterfactual profiles'' in certain theories. Take GRWm for example: when we evaluate a counterfactual situation by changing the mass densities within a bounded spatial region $R$, it is not at all clear how we should change the mass densities outside of $R$ or how we should change the wave function to reflect the changes in the mass densities as the mappings are many-to-one. There may be some ways to handle this worry (and we invite the defenders of 3N-Fundamentalism to address this worry), but the referee is right that it raises another sort of worry  on the viability of the functionalist program.}
\end{description}
More specifically, we take the 3N-Fundamentalist to be adopting the following sufficient condition:
\begin{description}
\item[Functionalism about pointers] If we have a fundamental theory of the world in which we can define a mapping from the fundamental degrees of freedom to $Q_1, Q_2, ... Q_n$, and $Q_1, Q_2, ... Q_n$ have the same counterfactual profile of what we take to be the 3-dimensional pointers in the Manifest Image, then $Q_1, Q_2, ... Q_n$ are real, emergent entities from that fundamental theory and our discourse  about pointer readings is made true by the fundamental theory via the functionalist mapping. 
\end{description}
Three questions arise. First, can we find such a mapping?  Answer: yes, it is available to the 3N-Fundamentalist. The most obvious solution would be to choose just the three degrees of freedom associated with each apparent particle from which we construct the configuration space. The 3-dimensional particles are not fundamental, but we recognize that there is a mapping from the 3N-dimensional space to N  particles in the 3-dimensional space. Therefore, our discourse about the pointers and human observers that are made out of particles in the 3-dimensional space is grounded via the functionalist mapping from the fundamental picture (in the Bohmian 3N picture: one Marvelous Point moving in a high-dimensional space).

Second, is the mapping unique?  Answer: it depends. We can, for example, use the dynamical structure to privilege one set of mappings.  Albert (1996) takes this strategy and suggests that there is a uniquely preferred mapping because of the contingent structure of the Hamiltonian of the universe that governs the dynamics of the fundamental degrees of freedom:

$$H=\sum_{i}^{\Omega} \frac{p_{i}^{2}}{2m_i}+\sum_{k, j = 1; k \neq j}^{\Omega/3} V_{j, k}([  (x_{3j-2} - x_{3k-2})^{2} + (x_{3j-1} - x_{3k-1})^{2} + (x_{3j} - x_{3k})^{2}  ])$$
The idea is that the fundamental degrees of freedom are contingently grouped into triples in the Hamiltonian potential term, instantiating a relation just like the familiar Pythagorean distance relation, which gives rise to the 3-dimensional Euclidean metric. Each triple is what we take to be a particle in the 3-dimensional space. Even in a classical world, the Pythagorean distance relation gives rise to a structure that is invariant under rotation, translation, and reflection symmetries, and grounds our claims about the 3-dimensional space. In a relativisitic world, the Minkowski metric, according to many people, gives rise to a genuinely 4-dimensional spacetime: $$d(a, b)^2=-(c\delta t)^2+(\delta x)^2+(\delta y)^2+(\delta z)^2$$ Therefore, the reasoning goes, the mathematical structure in the Hamiltonian potential term gives rise to an emergent, distinguished, and invariant structure of 3-dimensionality---$\mathbb{R}^{3}$---that grounds our discourse about pointer readings and human observers. Moreover, the Hamiltonian, in a unique way, puts the fundamental degrees of freedom into $\Omega/3$ triples.

Third, isn't the sufficient condition  too permissive? If what suffices to be emergent and real is just to stand in a mathematical mapping relation to the fundamental dynamical objects, then (as Tim Maudlin\footnote{Personal communication February 2015.} and John Hawthorne\footnote{See \cite{hawthorne2010metaphysician}, pp.147-152.}  observe independently) there will be many more emergent objects than we realize, a consequence that is highly implausible. Under \textbf{Functionalism}, simply by defining  a mathematical mapping from the location of every object in the world to three feet north of that object, there will be emergent objects  interacting with other emergent objects unbeknown to all of us! Since the 3-feet-north mapping is completely arbitrary, there will be infinitely many mappings, each of which will realize a different set of emergent entities. As a consequence, just by defining trivial mappings like these, we can create metaphysical monstrosities of an unimaginable scale. 

The 3-feet-north objection is a special case of a general objection to structuralism and functionalism: the mere existence of certain structure, counterfactual dependence, or causal relations is sometimes insufficient ground for establishing the reality of the emergent entities. In our context, the 3-feet-north objection suggests that the functionalist criterion is insufficient, and hence the proposed mapping from the fundamental degrees of freedom in 3N-Fundamentalism does not \emph{explain} the pointers or observers in the 3-dimensional space.  

I see two potential responses. First, the 3N-Fundamentalist can bite the bullet, embrace the seemingly absurd consequence of the functionalist criterion, and include in her ontology infinitely many sets of emergent entities. To show that the consequences are tolerable, she can give an argument that shows that all  sets of emergent entities are in fact equivalent in some way. One obvious argument makes use of relationalism about space. If relationalism is true, then there is no container (the substantial space) in addition to spatial relations among objects. Take any set of emergent entities (for example, the emergent entities that are 3 feet north of everything in the world), the spatial relations instantiated among them are the same spatial relations instantiated in any other set (for example, the emergent entities that are 5 feet west of everything in the world). The sets of emergent entities related by such mappings are just different ways of describing the same relational world. Call this \textbf{The Relationalist Approach}.

The Relationalist Approach, I think, is the most promising response on behalf of the 3N-Fundamentalist. However, it faces a problem. \textbf{Functionalism}, if true, seems to be a necessary truth. Relationalism, on the other hand, does not seem to be a necessary truth. The mere possibility of the failure of relationalism implies that there are possible worlds in which the functionalist criterion generates  too many emergent entities. Since those possible worlds are not too remote, our modal intuitions seem robust enough to account for them. The metaphysical monstrosity seems highly implausible. So it seems that \textbf{Functionalism} is still false. At this point, the 3N-Fundamentalist can reply that  \textbf{Functionalism} is in fact a contingent truth that holds only among relational worlds. It would take future work to show why this is true or how it follows from other commitments of \textbf{Functionalism}.

Instead of suggesting that the thesis is contingent, the 3N-Fundamentalist can reply that there are ways of restricting the functionalist criterion such that it does not lead to the absurd consequences. One way to restrict the criterion is to say that the functionalist mapping has to be an identity mapping.\footnote{Barry Loewer suggests this in personal communication, but he does not necessarily endorse it.} The reason that projection mappings from $\mathbb{R}^{3N}$ to $\mathbb{R}^{3}$ are insufficient is that  a projection map composed with a spatial translation is another a projection map; since the spatial translamtion can be completely arbitrary, there will be an infinite number of projection mappings that are equally good. However, if we let the functionalist relation to be the identity mapping, we can eliminate this problem. 
Applied to our case, the relation takes the triplets of degrees of freedom in the configuration space as \emph{identical} with the 3-dimensional particles. For example, take the Bohmian ``marvelous point'' in the 3N-dimensional configuration space: its coordinates $x_{1}, y_{1}, z_{1}$   just are a 3-dimensional particle. Identity is a strict relation that is not preserved by arbitrary mappings. Neither does the strategy rely on relationalism about space. Call this \textbf{The Identity Approach}.

However, not only does it have difficulty extending to GRW theories, the Identity Approach gets rid of the metaphysical monstrosities at the expense of eliminating the Manifest Image altogether. To see this, let us borrow some idioms from the grounding literature. The grounding relation, as commonly conceived,\footnote{For example, see \cite{RosenMDGR}.} is an umbrella term for several kinds of metaphysical dependence relations. One widely-accepted feature of grounding is that if a set of facts or entities, $\Sigma$, grounds another set of facts or entities $\Gamma$, then it is not the case that  $\Gamma$ grounds $\Sigma$. Such an asymmetry is a defining feature of the grounding relation. If the particles and their electromagnetic interactions ground the existence of a table, it is not the case that the table's existence grounds the particles and their electromagnetic interactions. Let us examine the suggestion that the functionalist mapping has to be an identity mapping. As I understand it, the functionalist mapping is a metaphysical dependence relation that falls under the umbrella of the grounding relation. Hence,  the  mapping has to be asymmetric. However, the identity  relation is symmetric. We have arrived at a contradiction. Therefore, one of the assumptions has to go. Since the functionalist mapping is supposed to metaphysically explain the emergence of 3-dimensional objects such as pointers and observers, it is best understood as a metaphysical explanation relation that is asymmetric. So it seems to me that the 3N-Fundamentalist should reject the suggestion that the functionalist criterion only allows identity mappings. 

There is another way to restrict the functionalist criterion. Instead of counting mathematical mappings of any sort, a 3N-Fundamentalist can restrict to  mappings between different spaces. For this smaller class of mathematical mappings, the 3-feet-north counterexamples do not arise, for the dynamics and causal behaviors on different levels are likely to be quite different. Call this \textbf{The Different Space Approach}. In addition to being obviously \emph{ad hoc} and unprincipled, the Different Space Approach does not block composite mappings. The composite mappings first from  $\mathbb{R}^{3N}$ to $\mathbb{R}^{3}$ then from $\mathbb{R}^{3}$ to itself are not ruled out by the restriction and still generate the same counterexamples. 

The above discussions of different approaches of functionalism---the Relationalist Approach, the Identity Approach, and the Different Space Approach---suggest that the current functionalist criterion is too permissive and leads to disastrous results for the 3N-Fundamentalists. Thus, we are right to doubt whether there can be any principled way to close the apparent explanatory gap in 3N-Fundamentalism. However, \textbf{Functionalism}, just like \textbf{Structuralism}, is still being developed, and its application here is novel and potentially promising. Given the recent work on the emergence of space-time and structural realism, there may well be future work that suggests better proposals than the ones we have considered. Anecdotal evidence suggests that many philosophers are not deterred by the above counterexamples and are still searching for some version of \textbf{Functionalism} that closes the explanatory gap in a principled way.


\bigskip


So far our discussions have focused on the two main considerations in the literature: the dynamical structure of the quantum theory and the successful explanation of the Manifest Image. If I am right, then the common arguments based on these considerations are either unsound or in need of refinement. After locating the crucial premises, I show how they can be resisted on either side. While the 3D-Fundamentalist can respond line by line by giving an alternative criterion of fundamental physical space, the 3N-Fundamentalist cannot provide a satisfactory functionalist criterion. So far, the considerations based on the dynamical structure and the Manifest Image are roughly in favor of 3D-Fundamentalism. However, anecdotal evidence suggests that this has not persuaded many 3N-Fundamentalists to switch sides, as they are often inclined to bite the bullet and swallow the costs of counter-intuitiveness. 

I take this to suggest that the most common philosophical arguments do not fully settle the debate between the two views. It looks like we have reached a stalemate. However, I suggest that we have not.  In the next section, by  looking into which view leads to a deeper understanding of the quantum world, I argue that we can find a new class of powerful arguments  that favor 3D-Fundamentalism over 3N-Fundamentalism. 


\section{Evidence \#3: Mathematical Symmetries in the Wave Function}

Having seen that the common arguments from the dynamical structure and ordinary experiences do not fully settle the debate between 3D-Fundamentalism  and 3N-Fundamentalism, I suggest we look at the debate from a different angle. As we are examining scientifically motivated metaphysical positions, we would like to see which one leads to a better or deeper understanding of the scientific phenomena. In this section, I offer an argument for 3D-Fundamentalism on the basis that it provides a deeper mathematical explanation of certain symmetries in the quantum theory. Since I do not take this to be the final word on the issue, I offer this as a first step toward an open research program to explore various mathematical explanations\footnote{Here I do not take sides whether such mathematical explanations are non-causal explanations.} in the quantum theories. 

\subsection{Another Argument for 3D-Fundamentalism}

\begin{description}
\item[P5] If a fundamental theory T explains S while T' does not, and S should be explained rather than postulated (other things being equal), then we have (defeasible) reasons to infer that T is  more likely than its alternative to be the fundamental theory.
\item[P6] 3D-Fundamentalism explains the Symmetrization Postulate but 3N-Fundamentalism does not. 
\item[P7] The Symmetrization Postulate should be explained rather than postulated. 
\bigskip\hrule\smallskip
\item[C3]  We have (defeasible) reasons to infer that 3D-Fundamentalism is more likely to be the fundamental theory than 3N-Fundamentalism. 
\end{description}

The first premise---\textbf{P5}---seems to me highly plausible. When we compare two fundamental theories, we measure them not just by empirical adequacy and internal coherence but also by explanatory depth. Say that a theory T provides a deeper explanation for X than T' does if T explains X from its axioms and T' postulates X as an  axiom. 

For example, if a fundamental physical theory T says that the \textbf{Symmetrization Postulate} is true (there are two groups of fundamental particles and one group has symmetric wave functions and the other group has anti-symmetric wave functions),  if (other things being equal) we would like to understand why that is the case, and if we find out that its alternative S provides an explanation for that fact, then we have defeasible reasons to favor S over T. 

\subsection{Justifying Premise 6}

If we accept \textbf{P5}, then the success of the argument depends on \textbf{P6} and \textbf{P7}. It is important to note that \textbf{P6} is not the unique premise that  delivers the conclusion. There probably are many mathematical explanations that favor 3D-Fundamentalism over 3N-Fundamentalism, including Lorentz symmetry.\footnote{For example, see \cite{allori2013primitive}, pp. 72-73.} (Indeed, I take it to be an open research question whether there are \emph{good} mathematical explanations that support 3N-Fundamentalism. To be sure, what counts as a good explanation and what counts as something that should be explained can be controversial.)

In any case, \textbf{P6} focuses on a particular mathematical explanation that arises from the mathematical study of identical particles and the nature of their configuration space.\footnote{In writing this section, I am grateful for many extensive discussions with Sheldon Goldstein and Roderich Tumulka about their papers on the Symmetrization Postulate.} 
\footnote{I do not include this explanation in the ``inferences from the dynamics'' category, for it concerns the mathematical construction of the theory.} 

Roughly speaking, identical particles share the same intrinsic physical properties, that is, the same qualitative properties recognized by the physical theory. For example, all electrons can be regarded as  identical (which in this context  means \emph{physically indistinguishable}) since they have the same  charge, mass, and spin. (To be sure, their different positions in space cannot correspond to their intrinsic properties, for if nothing else they would not be able to move.) Suppose we take the ordinary configuration space $\mathbb{R}^{3N}$ for N electrons. Then the configuration space is ordered: $$(x_1, y_1, z_1, x_2, y_2, z_2,..., x_n, y_n, z_n)$$
Each point in the configuration space is an ordered 3N-tuple and can be read as the $x,y,z$ coordinates of electron 1, those of electron 2, those of electron 3, and all the way up to the $x,y,z$ coordinates of electron N. The numeral labels  provide names (and thus distinguishability) to the electrons. Therefore, the configuration in which electron 1 and electron 2 are exchanged will be a different configuration and hence will be represented by a different point in the configuration space. 

However, the distinct points resulting from permutations of the electrons do not give rise to any real physical differences recognized by the physical laws. If we were to have the simplest ontology supporting the laws of quantum mechanics (or indeed any atomistic theory, including the Democritean atomism and classical mechanics), we should have a physical space representing all and only the real physical differences, excluding differences resulting from permutations of identical particles. 
Therefore, we can introduce the notion of an unordered configuration space for N  particles in 3-dimensional space as:

$$^{N}\mathbb{R}^{3} := \{S \subset R^{3} | \text{  cardinality}(S)=N\}$$
So a point in this configuration space would be: $$  \{(x_1, y_1, z_1), (x_2, y_2, z_2),..., (x_n, y_n, z_n)\} $$ Notice that the points are not 3N-tuples but sets of N elements, which are unordered (given the extensionality axiom of ZF set theory). In fact, it is a set of tuples, but the ordering corresponds to the coordinatization of the 3-dimensional space, which can be ultimately eliminated by using $\mathbb{E}^3$ instead of $\mathbb{R}^3$. 

Even in classical mechanics, we can use the unordered configuration space $^{N}\mathbb{R}^{3}$ (or the relevant unordered phase space) instead of the ordered configuration space $\mathbb{R}^{3N}$. Such a choice would not lead to any physical differences, as the two spaces share the same \emph{local} properties. They differ, however, in their  \emph{global} topological properties, which become essential in quantum mechanics. In particular, $^{N}\mathbb{R}^{3}$ (like the circle $S^1$) is topologically non-trivial in that it is not simply-connected, as not every closed loop is contractible to a point, while $\mathbb{R}^{3N}$ (like the real line $\mathbb{R}^{1}$) is simply connected---a much more trivial topology. As we will explain below, their global topological differences allow us to derive the deep and useful principle known as the Symmetrization Postulate.

Conventionally, a quantum theory with wave functions defined over an ordered configuration space needs to postulate an additional requirement---the Symmetrization Postulate---to rule out certain types of particles. It says that there are only two kinds of particles: symmetric wave functions for bosons and anti-symmetric wave functions for fermions. The ``symmetry'' and ``anti-symmetry'' refer to the behavior of the wave functions under position exchange. Stated for particles in 3-dimensions: 

\begin{description}
\item[Symmetrization Postulate:] There are only two kinds of wave functions: 
$$\text{(Bosons) } \qquad \psi_{B}(\textbf{x}_{\sigma (1)}, ..., \textbf{x}_{\sigma (N)}) = \psi_{B}(\textbf{x}_{1}, ..., \textbf{x}_{N}),$$  
$$\text{(Fermions) } \quad \psi_{F}(\textbf{x}_{\sigma (1)}, ..., \textbf{x}_{\sigma (N)}) = (-1)^{\sigma}\psi_{F}(\textbf{x}_{1}, ..., \textbf{x}_{N}),$$
where $\sigma$ is a  permutation of $\{1, 2, ..., N\}$ in the permutation group $S_N$, $(-1)^{\sigma}$ denotes the sign of $\sigma$, $\textbf{x}_{i} \in \mathbb{R}^{3}$ for $i = 1, 2, ..., N.$ 
\end{description}
This is an extremely deep and highly useful principle. To have a more intuitive grasp of it, we can recall the more familiar Pauli Exclusion Principle, which says that no two electrons can occupy the same quantum state (usually characterized by quantum numbers). For one direction of implication, consider two electrons (fermions) with a wave function that is totally anti-symmetric under position exchange. It follows that they cannot occupy the same position, that is: $$ \forall x_1, x_2 \in \mathbb{R}^3, \psi (x_1, x_2) = - \psi (x_2, x_1) \Longrightarrow \forall x  \in \mathbb{R}^3, \psi (x, x) = 0$$

Surprisingly, the Symmetrization Postulate, being deep and useful, emerges as a result of a beautiful mathematical analysis about the topological properties of the unordered configuration space.\footnote{For the classic papers in the mathematical physics literature, see \cite{dowker1972quantum} and \cite{leinaas1977theory}. But \cite{durr2006topological} and \cite{durr2007quantum} carry out the explanation much more thoroughly and successfully in Bohmian mechanics. The  outline in the next paragraph is a summary of their more technical derivation in the case of scalar wave functions.} Here we  sketch only an outline of the derivation; we refer interested readers to the Appendix for more technical details. 

Let us take Bohmian Mechanics (BM) as the background theory, in which there really are particles with precise trajectories, guided by a universal wave function. (Let us use scalar-valued wave functions to avoid further technicalities. We will later return to the question whether such an explanation is available on  the Copenhagen intepretation and GRW theories.) In BM, the universal wave function evolves according to the Schr\"odinger equation, and particles move according to the universal wave function and the guidance equation. The configuration space for N identical particles in $\mathbb{R}^{3}$ is $^{N}\mathbb{R}^{3}$, not $\mathbb{R}^{3N}$. However, they are intimately related. In fact, $\mathbb{R}^{3N}$ is what is called the ``universal covering space'' of $^{N}\mathbb{R}^{3}$. There is a natural projection mapping from  $\mathbb{R}^{3N}$ to $^{N}\mathbb{R}^{3}$ that forgets the ordering of particles. Since the physical configuration lies in $^{N}\mathbb{R}^{3}$, the velocity field needs to be defined there, whereas the wave function can still very well be defined on $\mathbb{R}^{3N}$, as we can project its velocity field from $\mathbb{R}^{3N}$ to the $^{N}\mathbb{R}^{3}$. However, not every velocity field can be so projected. To ensure that it can , we impose a natural ``periodicity condition'' on the wave function on $\mathbb{R}^{3N}$: 
$$\forall \hat{q} \in \mathbb{R}^{3N}, \sigma \in S_{N}, \psi(\sigma \hat{q})= \gamma_{\sigma} \psi(\hat{q}).$$ 
Since the fundamental group $S_N$ is a finite group, the topological factor $\gamma_{\sigma}$ has to be a group character (see \textbf{Unitarity Theorem} in Appendix). But $S_N$ has only two characters: (1) the trivial character $\gamma_{\sigma}=1$ and (2) the alternating character $\gamma_{\sigma}=sign(\sigma)=1 \text{ or } -1$ depending on whether $\sigma\in S_{N}$ is an even or an odd permutation. The former corresponds to the symmetric wave functions of bosons and the latter  to the anti-symmetric wave functions of fermions. Since any other topological factors are banned, we have ruled out other types of particles such as anyons. This result is equivalent to the Symmetrization Postulate.



We have seen that given some natural assumptions about identical particles in $\mathbb{R}^{3}$, the unordered configuration space $^{N}\mathbb{R}^{3}$, and a globally well-defined velocity field, we have very easily arrived at an explanation for the Symmetrization Postulate. With Bohmian Mechanics in the background, the derivation is well-motivated at each step. The rigorous mathematical work is done by Bohmian researchers in two beautiful papers \cite{durr2006topological} and \cite{durr2007quantum}, of which the above discussion is a simplification (of the simple case of scalar-valued wave functions). (For  technical details, please see the Appendix.) 

The above explanation might still work in the context of standard textbook quantum mechanics and the Copenhagen interpretation, according to which we should not take it seriously that particles are things that have precise locations and velocities simultaneously. Usually, in this context, we are told to act as if there are particles and to construct various mathematical objects from the degrees of freedom of the particles (such as the use of the configuration space). From a foundational point of view, it is not clear why the configuration space should be useful in such a theory. But if we play the usual game of acting \emph{as if} there are particles, perhaps we can also act \emph{as if} they are really identical,  \emph{as if} their states are given by a point in the unordered configuration space, and \emph{as if} the above assumptions in the Bohmian theory are also justified in this theory. Indeed, this seems to be the implicit attitude taken in two pioneer studies---\cite{dowker1972quantum} and \cite{leinaas1977theory}---about quantum mechanics for identical particles and unordered configuration space. 

What about other solutions to the measurement problem, such as GRWm (a spontaneous collapse theory with a mass-density ontology)? Since its fundamental material ontology consists in a mass-density field (not particles) in the 3-dimensional space, it is unclear why we should use the configuration space of particle locations. However, usually the wave function and the mass-density field are defined with the help of the configuration space. \footnote{For example, when defining the mass-density function at a time $t$, we start with a $|\Psi_{t}|^2$-distribution in the particle-location configuration space $\mathbb{R}^{3N}$, and obtain the value by using the marginal distribution weighted by the particle masses and summing over all particle degrees of freedom:  

$$\forall x\in \mathbb{R}^{3}, m(x,t) = \sum_{i=1}^{N} m_{i} \int_{\mathbb{R}^{3N}} d^{3} x_{1} ... d^{3} x_{N} \delta^{3} (x_{i} - x) |\Psi_{t}(x_{1},..., x_{N})|^2  $$}
Nevertheless, it seems inconsistent with the spirit of GRWm to act as if there are fundamental particles or as if the above-mentioned Bohmian assumptions hold in the theory. In particular, there do not seem to be any compelling reasons to consider $^{N}\mathbb{R}^{3}$, a crucial piece in the above derivation, unlike in the case of BM. Hence, we do not think that the previous argument applies smoothly to the case of GRW theories or the Many-Worlds interpretation with a mass-density ontology.\footnote{We thank an anonymous referee for helping us make this clear.} Unlike the arguments in the previous sections, the argument we offer here does depend on the specific interpretation of quantum mechanics, namely Bohmian Mechanics. Whether we can make similar topological arguments in the context of GRW and MWI would require thinking more deeply about the structures of their state spaces. (This stands in contrast with the situation in \S 1 and \S 2, where the arguments do not require a justification of the structure of the configuration space.)

Returning to the construction of the unordered configuration space $^{N}\mathbb{R}^{3}$, we observe that it stands in a special relation to $\mathbb{R}^{3}$, namely, mathematical construction. In this sense, a quantum system of N-identical-particles in $\mathbb{R}^{3}$ naturally give rise to $^{N}\mathbb{R}^{3}$, which in turn naturally give rise to the mathematical explanation of the Symmetrization Postulate. 

However, this explanation is not available for the quantum theory on 3N-Fundamentalism. Although in the above explanation we use $\mathbb{R}^{3N}$ in defining the wave function (as it is the universal covering space of $^{N}\mathbb{R}^{3}$), simply starting with $\mathbb{R}^{3N}$ gets us nowhere near to the characters of the permutation group. Starting with a wave function and a ``marvelous point'' (the only fundamental particle) on $\mathbb{R}^{3N}$, we do not have any motivation to consider $^{N}\mathbb{R}^{3}$, the natural configuration space for $N$ particles in $\mathbb{R}^{3}$. Consequently, we cannot make use of its explanatory power in the derivation of the Symmetrization Postulate. (That is because, given a particle moving in $\mathbb{R}^{3N}$, we do not have any motivation to use the covering-space construction to derive the topological factors of the wave function, and we do not have an explanation for the Symmetrization Postulate.)


Therefore,  we have justified \textbf{P6}---the second premise in the argument. 

\subsection{Deep versus Shallow Explanations}

What about \textbf{P7}, the idea that the Symmetrization Postulate should be explained rather than postulated?\footnote{\textbf{P7} shows the difference between my argument here and Maudlin's argument in \cite{maudlin2013nature} pp.140-42. The difference in our arguments lies in our use of different explananda. Maudlin suggests that for, say,  an 8-particle universe, the natural configuration space is $^{8}\mathbb{R}^{3}$ instead of $\mathbb{R}^{24}$, but that cries out for an explanation: why is this particular space $^{8}\mathbb{R}^{3}$ the fundamental space instead of some other space with a different factorization of 24 (such as $^{4}\mathbb{R}^{6}$)? This is an insightful argument, but I do not think it would \emph{convince} many defenders of  3N-Fundamentalism, as on their view the fundamental space does not  need  further explanation. In contrast, our argument is quite different: we are not asking for an explanation of the fundamental space, but for an explanation of the Symmetrization Postulate, something that everyone agrees to be a deep and puzzling symmetry in quantum mechanics. Thus, I include Pauli's quote as it beautifully conveys the deep puzzlement. } 
Here are some compelling reasons to endorse it:

\begin{itemize}
\item Its form is very different from the other laws of QM such as the Schr\"{o}dinger equation. It is a direct restriction on the possible quantum states of systems with indistinguishable particles. 
\item It has high explanatory power. The Symmetrization Postulate explains Pauli's Exclusion Principle and the Periodic Table. It also explains why there are exactly two kinds of particles in QM.
\item In the words of Pauli, at the occasion of his 1946 Nobel Lecture, as quoted in the epigraph: ``Already in my original paper I stressed the circumstance that I was unable to give a logical reason for the exclusion principle or to deduce it from more general assumptions. I had always the feeling, and I still have it today, that this is a deficiency.''\footnote{\cite{pauli1994writings}, p.171.}
\end{itemize}
Granted, one might still deny \textbf{P7} at this point. But it seems to me (and to many working physicists) that \textbf{P7} is much more likely to be true than its denial. The above reasons suggest that an explanation of \textbf{P7} would point us to something deep in the physical nature of the world. It is true that in an ultimately axiomatic theory of physics, the explanatory relations can be reversed (just as logical theories can be axiomatized in different ways corresponding to different explanatory relations). However, I think it is reasonable to believe that the final quantum theory ought not to postulate the Symmetrization Postulate as fundamental but rather as something that follows from simpler axioms. 


\subsection{What about $^{N}\mathbb{R}^{3}$-Fundamentalism?}

In this essay, I have assumed that the 3N-Fundamentalist is really a $\mathbb{R}^{3N}$-Fundamentalist. She can of course modify her thesis and believe instead in a fundamentally multiply-connected physical space, $^{N}\mathbb{R}^{3}$. The 3D-Fundamentalist can respond by charging her opponent's move as \emph{ad hoc} or unnatural, since $^{N}\mathbb{R}^{3}$ is ``clearly'' about $N$ identical particles in $\mathbb{R}^{3}$. But similar criticisms have not prevented anyone from taking $\mathbb{R}^{3N}$ as fundamental, even though in some sense it is ``clearly'' about $N$ particles in $\mathbb{R}^{3}$. 

\begin{center}
\begin{tabular}{|c|c|}
\hline
3D-Fundamentalism & 3N-Fundamentalism \\ \hline \hline
$\mathbb{R}^{3}$+distinguishable particles &  \\ 
 & $\mathbb{R}^{3N}$ \\ \cline{1-1}
 & \\ \cline{2-2}
$\mathbb{R}^{3}$+indistinguishable particles & $^{N}\mathbb{R}^{3}$ \\ \hline
\end{tabular}
\end{center}

There is, however, another argument against the 3N-Fundamentalist's maneuver that is even less controversial. My main goal in this paper has been to evaluate the evidential support for 3D-Fundamentalism and 3N-Fundamentalism. Given 3N-Fundamentalism, the fundamental physical space has two main possibilities: (1) a simply-connected space $\mathbb{R}^{3N}$ and (2) a multiply-connected space that is a quotient space of $\mathbb{R}^{3N}$ by the permutation group of $k$ objects where $k$ divides $3N$. I believe that (1) is intrinsically more likely than (2). But suppose we grant them equal probability. Still, (2) comes in many further possibilities, and $^{N}\mathbb{R}^{3}$ is intrinsically as likely as any other quotient space. So we should assign equal probability to each possibility in (2). Given a large $N$, $^{N}\mathbb{R}^{3}$ will receive a very small probability compared to that of $\mathbb{R}^{3N}$.\footnote{Thanks to Gordon Belot for helpful discussions here.} In the table above, the ratio of the two areas for $\mathbb{R}^{3N}$ and $^{N}\mathbb{R}^{3}$ roughly corresponds to the relative confidence that we have in them. 

Given 3D-Fundamentalism, there is only one main candidate for the fundamental physical space--$\mathbb{R}^{3}$ and two main candidates for the configuration space--$\mathbb{R}^{3N}$ and $^{N}\mathbb{R}^{3}$. Although an ontology of identical particles is simpler than that of non-identical particles, for the sake of the argument, we can give these two possibilities equal weight. Hence, in the table above, the two candidates divide the area roughly in half. 

If we take as a datum that  the Symmetrization Postulate is true,  assume that $^{N}\mathbb{R}^{3}$ is the correct route to explain it, and we allow explanatory strength to come in degrees, then overall (considering all versions) 3D-Fundamentalism explains it better than 3N-Fundamentalism does. So although both 3D-Fundamentalism and 3N-Fundamentalism contain versions of them that explain the Symmetrization Postulate, 3D-Fundamentalism is better supported by the successful explanation. Therefore, the evidence ``disconfirms'' 3N-Fundamentalism over 3D-Fundamentalism. (That is, if we assume a standard way of thinking about update and Bayesian confirmation theory.\footnote{In a different context (the problem of evil) at the Rutgers Religious Epistemology Workshop in May 2014, Lara Buchak discussed a different approach to rational updating---updating by conditionals. It is up to the defender of 3N-Fundamentalism to develop and apply that approach here. Even if that were to succeed, however, I think it would be a significant concession that 3N-Fundametnalism is disconfirmed in the ``Bayesian'' sense.}) In fact, we can reformulate the original argument by replacing \textbf{P6} with a weaker premise:

\begin{description}
\item[P6*] 3D-Fundamentalism explains the Symmetrization Postulate significantly better than 3N-Fundamentalism. 
\end{description}
Therefore, even if $^{N}\mathbb{R}^{3}$-Fundamentalism is an option for the 3N-Fundamentalism, her view is still not well supported by the explanation. The above argument fleshes out our intuition that the 3N-Fundamentalist maneuver is \emph{ad hoc} or unnatural.



However, this should reflect only our current state of knowledge and should not be taken as the final word on the issue. Indeed, there are several ways that future research can go:
\begin{enumerate}
\item We discover more mathematical explanations only available to 3D-Fundamentalism, which gives us more reason to favor 3D-Fundamentalism over 3N-Fundamentalism. 
\item We discover some mathematical explanations only available to 3N-Fundamentalism, which restores or reverses the evidential balance between 3D-Fundamentalism and 3N-Fundamentalism. 
\item We discover that there exists a mathematical explanation of the Symmetrization Postulate from other resources in 3N-Fundamentalism that is independent from the topological considerations as discussed above,  which restores the evidential balance between 3D-Fundamentalism and 3N-Fundamentalism. 
\end{enumerate}

\begin{center}
\begin{table}
\begin{tabular}{l | c | c | c | c }
Considerations & 3D-Fundamentalism & 3N-Fundamentalism \\
\hline \hline
Dynamics & 0/+ & + \\ 
Manifest Image & + & -/0 \\
Explanatory Depth & + & -/0\\
\end{tabular}
\caption{Summary of pros and cons; where ``+'' means ``strongly in favor,'' ``-'' means ``strongly against,'' and ``0'' means ``roughly neutral.''}
\end{table}
\end{center}

\section{Conclusion}

Based on the above evaluation of the three kinds of evidence (see Table 1), I conclude that, given our current knowledge, it is more likely that the fundamental physical space in quantum mechanics is 3-dimensional rather than 3N-dimensional. However, as noted above, there are future directions of research where the 3N-Fundamentalists can restore or even reverse the evidential balance.

Our debate here is related to the discussions about theoretical equivalence; we observe the following:

\begin{enumerate}
\item 3N-Fundamentalism and 3D-Fundamentalism are highly equivalent in theoretical structure--the mathematics used is exactly the same; 
\item The debate between 3N-Fundamentalists and 3D-Fundamentalists almost reaches a stalemate and hence might be considered as non-substantive;
\item An important new piece of evidence lies not in the inferences from the dynamics or ordinary experiences but in the mathematical symmetries in the wave function. What breaks the tie is the fact that 3N-Fundamentalism and 3D-Fundamentalism are explanatorily inequivalent, and the latter explains the Symmetrization Postulate better than the former. 
\end{enumerate}

Therefore, the discussion about the wave function provides another useful case for the ongoing debate about theoretical equivalence and structure. The connection should prove fruitful for future research.\footnote{Thanks to David Schroeren for discussing this last point with me.}

\section{Appendix: A Topological Explanation of the Symmetrization Postulate}

Here, we provide the technical details of our argument in \S 3.2 by following the mathematical derivations in \cite{durr2006topological} and \cite{durr2007quantum} regarding the case of scalar-valued wave functions. We omit the case of vector-valued wave functions as it is mathematically more complicated but conceptually similar to the scalar case.

Our goal is to arrive at all possible Bohmian dynamics for $N$ identical particles moving in $\mathbb{R}^3$, guided by a scalar wave function. We want to show that, given some natural assumptions, there are only two possible dynamics, corresponding to the symmetric wave functions for bosons and the anti-symmetric wave functions for fermions. We will use the covering-space construction and examine the permissible topological factors.\footnote{This argument, of course, is not intended as a mathematical demonstration, as we appeal to considerations of naturalness and simplicity. But these assumptions are relatively weak, and they are guided by strong intuitions of the practicing mathematical physicists. Even better, in our case of $^{N}\mathbb{R}^{3}$, we can prove the \textbf{Unitarity Theorem}---that the topological factor in the periodicity condition has to be a character of the fundamental group.}

The natural configuration space for $N$ identical particles in $\mathbb{R}^3$ is:

$$^{N}\mathbb{R}^{3} := \{S \subset R^{3} | \text{  cardinality}(S)=N\}$$
Since it is a multiply-connected topological space, we follow the usual covering-space construction to define the dynamics on its universal covering space and project it down to $^{N}\mathbb{R}^{3}$. Its universal covering space is, unsurprisingly, the simply-connected  $\mathbb{R}^{3N}$. Let $Cov(\mathbb{R}^{3N}, ^{N}\mathbb{R}^{3})$ denote the covering group of the base space $ ^{N}\mathbb{R}^{3}$ and its universal covering space $\mathbb{R}^{3N}$. Given a map $\gamma: Cov(\mathbb{R}^{3N}, ^{N}\mathbb{R}^{3}) \rightarrow \mathbb{C}$, we define a wave function on this space: $\psi: \mathbb{R}^{3N} \rightarrow \mathbb{C}$ with the following periodicity condition associated with $\gamma$:
$$\forall \hat{q} \in \mathbb{R}^{3N}, \forall \sigma \in Cov(\mathbb{R}^{3N}, ^{N}\mathbb{R}^{3}), \psi(\sigma \hat{q})= \gamma_{\sigma} \psi(\hat{q}).$$

First, we show that if the wave function does not identically vanish (which is a natural assumption), the topological factor $\gamma$ is a representation. Given any $\sigma_{1}, \sigma_{1} \in Cov(\mathbb{R}^{3N}, ^{N}\mathbb{R}^{3})$:

$$\gamma_{\sigma_{1}\sigma_{2}} \psi(\hat{q}) = \psi(\sigma_{1}\sigma_{2}\hat{q})= \gamma_{\sigma_{1}} \psi(\sigma_{2}\hat{q}) = \gamma_{\sigma_{1}} \gamma_{\sigma_{2}}\psi(\hat{q}). $$
Hence, $\gamma_{\sigma_{1}\sigma_{2}} = \gamma_{\sigma_{1}}\gamma_{\sigma_{2}}.$ Therefore, $\gamma$ is a structure-preserving map and a representation of the covering group. 

It is a well-known fact that the covering group is canonically isomorphic to the fundamental group
 $\pi_{1} (Q, q)$, where $Q$ is the base space. In our case, the fundamental group  $\pi_{1} (^{N}\mathbb{R}^{3}, q)$  is $S_{N}$, the group of permutations of $N$ objects. It has only two characters:   (1) the trivial character $\gamma_{\sigma}=1$ and (2) the alternating character $\gamma_{\sigma}=sign(\sigma)=1 \text{ or } -1$ depending on whether $\sigma\in S_{N}$ is an even or an odd permutation. The former corresponds to the symmetric wave functions of bosons and the latter  to the anti-symmetric wave functions of fermions. If we can justify the use of the periodicity condition and a ban on any other topological factors in the periodicity condition, we would be able to rule out other types of particles such as anyons. This result would be equivalent to the Symmetrization Postulate.  

The periodicity condition is the most natural condition to require if we want a projectable velocity field given by the wave function on the covering space. The periodicity condition implies that $\nabla \psi(\sigma \hat{q}) = \gamma_{\sigma} \sigma^{*} \nabla \psi(\hat{q}),$ where $\sigma^{*} $ is the push-forward action of $\sigma$ on tangent vectors. Hence, if we define the velocity field on $\mathbb{R}^{3N}$ in the usual way: 
$$\hat{v}^{\psi}(\hat{q}) := \hbar \text{Im} \frac{\nabla \psi}{\psi} (\hat{q}),$$
then it is related to other levels of the covering space:
$$\hat{v}^{\psi}(\sigma\hat{q}) = \sigma^{*}\hat{v}^{\psi}(\hat{q}).$$
This makes $\hat{v} $ projectable to a well-defined velocity field on the base space $^{N}\mathbb{R}^{3}$:
$$v^{\psi}(q) = \pi^{*}\hat{v}^{\psi}(\hat{q}),$$
where $\hat{q}$ is an arbitrary point in $\mathbb{R}^{3N}$ that projects down to $q$ via the projection map $\pi$.

Moreover, the periodicity condition is natural because  it is preserved by the Schr\"odinger evolution. Therefore, given an initial point $q_0$ in the unordered configuration space $^{N}\mathbb{R}^{3}$, we can choose an arbitrary point $\hat{q_0}$ in the ordered configuration space $\mathbb{R}^{3N}$ that projects to $q_0$ via $\pi$, evolve $\hat{q_0}$ by the usual guidance equation in $\mathbb{R}^{3N}$ until time $t$, and get the final point $q_t=\pi(\hat{q_{t}})$. The final point $q_t \in  ^{N}\mathbb{R}^{3}$ is independent of the choice of the initial $\hat{q_0}\in \mathbb{R}^{3N}$.


We explain why it is crucial for the topological factor to be not only a representation of the fundamental group but also unitary, as it is required to ensure that the probability distribution is equivariant. Given the periodicity condition, we have 
$$|\psi(\sigma\hat{q})|^{2} = |\gamma_{\sigma}|^{2} |\psi(\hat{q})|^{2}. $$
To make $|\psi(\hat{q})|^{2} $ projectable to a function on $ ^{N}\mathbb{R}^{3}$, we require that $\forall \sigma\in S_{N}, |\gamma_{\sigma}|^{2}=1$. This is equivalent to $\gamma$ being a character (a unitary representation) for the fundamental group. Given the Schr\"odinger equation on $\mathbb{R}^{3N}$ and the projection of $|\psi(\hat{q})|^{2} $ to $|\psi|^{2} (q),$ we have:
$$\frac{\partial |\psi_{t}|^{2} (q)}{\partial t} = - \nabla \cdot (|\psi_{t}|^{2} (q) v^{\psi_{t}}(q)). $$
Compare this with the transport equation for a probability density $\rho$ on $^{N}\mathbb{R}^{3}$:
$$\frac{\partial \rho_{t} (q)}{\partial t} = - \nabla \cdot (\rho_{t} (q) v^{\psi_{t}}(q)). $$
Therefore, if $\rho_{t_0}(q) = |\psi_{t_0}|^{2} (q)$ at the initial time $t_0$, then $\rho_{t}(q) = |\psi_{t}|^{2} (q)$ at all later times. We have arrived at equivariance. 

The above argument works for the general case of multiply-connected spaces as well as the particular case of $^{N}\mathbb{R}^{3}$. In our case of $^{N}\mathbb{R}^{3}$, we can prove the following simple theorem that the topological factor $\gamma$ must be a unitary representation, i.e. a group character. 

\newtheorem{innercustomthm}{Unitarity Theorem}
\newenvironment{customthm}[1]
  {\renewcommand\theinnercustomthm{#1}\innercustomthm}
  {\endinnercustomthm}
\newcommand*{\QEDB}{\hfill\ensuremath{\square}}%

\begin{customthm}{}
Let $\sigma \in Cov(\mathbb{R}^{3N}, ^{N}\mathbb{R}^{3})$,  $ \gamma: Cov(\mathbb{R}^{3N}, ^{N}\mathbb{R}^{3}) \rightarrow \mathbb{C}$. If $\gamma_{\sigma}\neq 0$ and $\gamma_{\sigma_{1}\sigma_{2}} = \gamma_{\sigma_{1}}\gamma_{\sigma_{2}},$
 $\forall \sigma_{1}, \sigma_{2} \in Cov(\mathbb{R}^{3N}, ^{N}\mathbb{R}^{3})$, then $|\gamma_{\sigma}|=1$.

\end{customthm}

\begin{proof}
The fundamental group of $^{N}\mathbb{R}^{3}$ is the permutation group $S_N$, which has $N!$ elements. We obtain that $\gamma_{Id}=1$, because $\gamma_{\sigma}=\gamma_{(\sigma \ast Id)}=\gamma_{\sigma}\gamma_{Id}$ and $ \gamma_{\sigma}\neq 0$. It is a general fact that in a finite group with $k$ elements, every element $\sigma$ satisfies $\sigma^k=Id.$ Therefore, rewriting $\gamma_{\sigma}=Re^{i\theta}$, we have $1e^{0i}=1=\gamma_{Id}=\gamma_{\sigma^k}=(\gamma_{\sigma})^{k}=R^{k}e^{ik\theta}$. So we have: $|\gamma_{\sigma}|=1$, which makes $\gamma_{\sigma}$ a unitary representation of the covering group, which is a character of the fundamental group. \QEDB
\end{proof}

Therefore, the periodicity condition associated with the topological factor:
$$\forall \hat{q} \in \mathbb{R}^{3N}, \forall \sigma \in S_{N}, \psi(\sigma \hat{q})= \gamma_{\sigma} \psi(\hat{q})$$
is a highly natural and simple condition that guarantees well-defined dynamics on $^{N}\mathbb{R}^{3}$, and the topological factors are the characters of the fundamental group.\footnote{In our particular case of $^{N}\mathbb{R}^{3}$, we have the \textbf{Unitarity Theorem} to explain why the topological factors have to be  characters of the fundamental group. In the general case, even without such a theorem there are still many good reasons why we should restrict the topological factors to the group characters.  
See \cite{durr2007quantum}  \S 9 ``The Character Quantization Principle.''} In the case of $^{N}\mathbb{R}^{3}$, the fundamental group is the permutation group of $N$ objects: $S_{N}$. Recall that it has only two characters: (1) the trivial character $\gamma_{\sigma}=1$ and (2) the alternating character $\gamma_{\sigma}=sign(\sigma)=1 \text{ or } -1$ depending on whether $\sigma\in S_{N}$ is an even or an odd permutation. This leads to two possible dynamics corresponding to the symmetric and the anti-symmetric wave functions (and no more): 

$$\text{(Bosons) } \qquad \psi_{B}(\textbf{x}_{\sigma (1)}, ..., \textbf{x}_{\sigma (N)}) = \psi_{B}(\textbf{x}_{1}, ..., \textbf{x}_{N}),$$  
$$\text{(Fermions) } \quad \psi_{F}(\textbf{x}_{\sigma (1)}, ..., \textbf{x}_{\sigma (N)}) = (-1)^{\sigma}\psi_{F}(\textbf{x}_{1}, ..., \textbf{x}_{N}),$$
where $\sigma$ is a  permutation of $\{1, 2, ..., N\}$ in the permutation group $S_N$, $(-1)^{\sigma}$ denotes the sign of $\sigma$, $\textbf{x}_{i} \in \mathbb{R}^{3}$ for $i = 1, 2, ..., N.$ Therefore, we have arrived at the statement of the Symmetrization Postulate for the 3-dimensional physical space. Interestingly, the same argument would predict that there are more possibilities than fermions and bosons if we go to a smaller space. For $N$ identical particles in a 2-dimensional space, there is the additional possibility of fractional statistics, corresponding to anyons.\footnote{Perhaps this is yet another reason why the 3-dimensionality of the physical space is a distinguished feature of reality.}

\section*{Acknowledgement}
I am grateful for insightful comments from the anonymous reviewers at \emph{The Journal of Philosophy} that  led to not only significant improvements throughout the paper but also the use of Pauli's quote in the epigraph, and for helpful discussions with David Albert, Barry Loewer, Roderich Tumulka, Sheldon Goldstein, Nino Zangh\`i, Tim Maudlin, Gordon Belot, Jonathan Schaffer, Michael Townsen Hicks, Jill North, Isaac Wilhelm, Thad Roberts, Bradley Monton, audiences at the Rutgers Metaphysics Reading Group, the 2015 Black Forest Summer School in Philosophy of Physics (Saig), a joint meeting with the Beyond Spacetime Project at University of Illinois--Chicago and University of Geneva, the 2015 Annual Meeting of the Society for the Metaphysics of Science (Newark), where Vishnya Maudlin provided  insightful comments that led to much rethinking about \S 1, and a symposium discussion at the 2016 Central APA Meeting (Chicago), where Nick Huggett provided eight pages of helpful comments that led to many improvements in \S 2 and \S 3. 

\bibliography{test}


\end{document}